\documentstyle[preprint,aps]{revtex}
\begin{document}
\draft
\preprint{\hfill\vbox{\hbox{IUCAA-8/98}}}
\title{\bf Naked singularities in low energy, effective string theory 
}
\author{Sayan Kar \thanks{Electronic Address :
sayan@iucaa.ernet.in}} 
\address{Inter University Centre for Astronomy and Astrophysics,\\
Post Bag 4, Ganeshkhind, Pune, 411 007, INDIA}
\maketitle
\parshape=1 0.75in 5.5in
\begin{abstract}
Solutions to the equations of motion of the low energy, effective field theory
emerging out of compactified heterotic
string theory are constructed by making
use of the well--known duality symmetries. Beginning 
with four--dimensional solutions of the Einstein--massless
scalar field theory in the canonical frame we first rewrite the
corresponding solutions in the string frame. Thereafter, using the
T and S duality symmetries of the low energy string effective action   
we arrive at the corresponding uncharged, electrically charged and 
magnetically charged solutions. Brief comments on the construction
of dual versions of the Kerr-Sen type using the
dilatonic Kerr solution as the seed are also included. Thereafter,
we verify the status of the
energy conditions for the solutions in the string frame. Several
of the metrics found here are shown to possess naked singularities
although the energy conditions are obeyed. Dual solutions exhibit
a duality in the conservation/violation of the Null and Averaged
Null Energy Conditions (NEC/ANEC), a fact demonstrated earlier in the
context of black holes (hep-th/9604047) and cosmologies
(hep-th/9611122).
Additionally, those backgrounds which conserve the NEC/ANEC 
in spite of possesing naked singularities serve as 
counterexamples to cosmic censorship in the
context of low energy, effective string theory.  
\end{abstract}

\vskip 0.125 in
\parshape=1 0.75in 5.5in


\newpage

\section{\bf Introduction}
 
The equations of motion for the background fields (metric, dilaton
and the antisymmetric tensor field) of string theory, 
obtained by setting
the $\beta$-functions of the $\sigma$- model to zero, are known to
have solutions 
representing
black holes {\cite{bh:str}}, cosmologies {\cite{cos:str}}
etc.. Extensive investigations
about various features of these geometries have been carried out over
the last decade or so. It may therefore seem somewhat surprising that there
are still newer solutions with distinguishing characteristics. The fact
that the solution--set has not been completely exhausted is largely
due to the nonlinearity of the equations (in the same way as in General
Relativity(GR))
as well as the large number of extra matter fields that arise in compactified
, low--energy effective string theory. Therefore, it would not
be improper to state that there still does exist a fair amount 
of scope so far as solution--construction is concerned. 

In this paper, we first revisit the 
well--known solutions of the Einstein--scalar field system which
represent naked singularities. Among these spacetimes are the
ones constructed by Janis, Newman and Winicour {\cite{jnw:prl68}}
which have been recently shown to be equivalent to the 
Wyman solutions {\cite{wy:81}} in {\cite{ksv:ijmp97}}. 
There are further generalisations of these solutions for the
non--static case discussed by Roberts {\cite{rob:grg89}} and more
recently in {\cite{hmn:prd94}} in the context of scalar
field collapse. Higher dimensional analogs of the
Wyman solutions have been constructed in {\cite{xz:prd89}}
with essentially the same features (singular event horizons).
All these metrics when written in the string/Einstein
frame are solutions of the background field equations in the
corresponding frame.
Using these geometries as {\em seed} solutions, we construct 
newer examples by exploiting the symmetries of low energy,
effective string theory. In particular, we first use T--duality
{\cite{td:ref} for the string frame metric to
construct the electrically charged solutions. Then, we use the S-duality
symmetry {\cite{sen:swc}} to
write down the magnetic counterpart of the electric solution in the
string frame. We also use other kinds of transformations--basically
subsets of the full T--duality group, such as the Buscher formulae
{\cite{bush:plb}} to
find examples in a different class. 

It should be mentioned that in a recent paper {\cite{kp:npb97}} 
Kiem and Park have obtained general solutions of the 
Einstein--Maxwell--dilaton theory in $D$ dimensions.
These authors exploit the fact that the D dimensional theory
can be reduced to an effective two dimensional dilaton
gravity model by a spherisymmetric ansatz for the 
$D$ dimensional metric. Some of the solutions in their
paper do already exist in the literature (notably the
purely dilatonic solutions in diverse dimensions). Our first aim
here is to demonstrate how, using duality properties one
can indeed obtain generalisations which do reduce
to the known solutions under specific choices of certain parameters.

The $new$ solutions in this paper are : (a) the magnetically
charged solution in the string frame and its various limiting
cases (b) the $T$ dual pure scalar field solution in the string
frame obtained using the Buscher formulae (c) the electrically
charged solution in the string frame obtained by using an 
$O(2,1)$ transformation of the string frame JNW solution.
Additionally, we chart out how the choice of the four
different parameters (namely $Q,M,\sigma,\alpha$ ) gives
rise to the various geometries, some of which are already
known in the literature.   

After arriving at the various solutions we then move on towards
constructing their physical properties and, more importantly,
 also investigate the  
the nature of the singularities and  the energy conditions. 
Most of the metrics we obtain actually possess naked singularities
even though they seem to obey the energy conditions. This prompts
us to comment that within the context of low energy, effective, string theory 
we have
examples of violations of cosmic censorship. This fact was noted
for the Einstein--scalar system in the Einstein frame 
by Roberts and there do exist
recent verifications of this in the context of scalar field
collapse through the numerical work of
Choptuik and others {\cite{matt:prl93}}
 which indicate the existence of point--mass
black holes (essentially naked singularities!). Finally, we demonstrate
a reflection of string dualities in the conservation/violation of the
local and global null energy condition--a fact which was noticed
earlier for black holes and cosmologies in earlier papers by
this author {\cite{sk:prd97}}, {\cite{sk:plb97}}.  

It may be argued that there is no need to look into the
properties of matter in the {\em string frame} because the
Einstein frame solutions do not have any problem with
the energy conditions (recall that the dilaton
kinetic term has the right sign in the action written in
the canonical Einstein frame). However, it is an acceptable fact that
no consensus has been reached as yet about which frame is
{\em more} important in string theory. This is largely due
to the validity of certain symmetries in
the string frame (eg. T--duality) whereas others hold good in the 
Einstein frame (eg. S--duality).
Moreover, since the string frame metric is the one which appears in
the nonlinear $\sigma$--model, whose $\beta$ functions we set to
zero (in order to implement quantum conformal invariance) to arrive at the
low energy effective field theory
it is this metric that a `string sees'.  

The paper is organised as follows. We first write down the relevant
actions and the equations of motion in II and also the known solutions
of the Einstein--massless--scalar system. Then, using the duality
transformations we enumerate the various charged as well as 
uncharged solutions. In Section III the properties of the various
solutions are discussed--the singularities as well as the 
energy conditions. Section IV contains a summary and concluding
remarks. The Appendix to the paper contains relevant expressions
for the Riemann. Ricci, Einstein tensors and the Ricci scalar for the
class of metrics under consideration. 

We follow the sign conventions of MTW {\cite{mtw:73}}.  

\section{\bf Action, equations and solutions}

Our starting point is the four dimensional,
 low energy effective action of heterotic string
theory compactified on a six--torus. The field content is 
--the graviton ($g_{\mu\nu}$), antisymmetric tensor field ($B_{\mu\nu}$),
Maxwell field, ($A_{\mu}$), dilaton ($\phi$). We have retained only
one of the 16 U(1) gauge fields present in the action written in, say
{\cite{sen:swc}} and also set the various moduli fields (M) 
to zero. 

The action is given as :

\begin{equation}
S_{eff} = \int d^{4}x \sqrt{-g} e^{-2\phi} \left [ R + 4\left (\nabla \phi
\right ) ^{2} - \frac{1}{12} H_{\mu\nu\rho}H^{\mu\nu\rho} -
F_{\mu\nu}F^{\mu \nu} \right ] 
\end{equation}

The corresponding equations of motion obtained by performing 
variations with respect to the various fields $g_{\mu\nu}$, $B_{\mu\nu}$,
$\phi$, and $A_{\mu}$ are given as follows :

\begin{equation}
R_{\mu\nu} = -2\nabla_{\mu}\nabla_{\nu} \phi + 2F_{\mu\lambda}F^{\lambda}
_{\nu} + \frac{1}{4}H_{\mu\lambda\sigma}H_{\nu}^{\lambda\sigma} 
\end{equation} 

\begin{equation}
\nabla^{\nu} \left ( e^{-2\phi}F_{\mu\nu} \right ) + \frac{1}{12}e^{-2\phi}
H_{\mu\nu\rho}F^{\nu\rho} = 0
\end{equation}

\begin{equation}
\nabla^{\mu} \left (e^{-2\phi} H_{\mu\nu\rho} \right ) = 0
\end{equation}

\begin{equation}
4\nabla^{2} \phi - 4\left (\nabla \phi \right ) ^{2} + R - F^{2} -
\frac{1}{12}H^{2} = 0
\end{equation}

The above set of equations are written for the metric in the
string (Brans--Dicke) frame.

Assuming the Maxwell and antisymmetric tensor fields to be zero
(note that the zero values are consistent with the equations
for these fields and do not impose additional restrictions on
the metric or the dilaton)  we 
get the following spherically symmetric, static solution :

\begin{equation}
ds_{str.}^{2} = -\left ( 1 - \frac{2\eta}{r} \right )^\frac{m+ \sigma}{\eta}
dt^{2} + \left ( 1 - \frac{2\eta}{r} \right )^{\frac{\sigma - m}{\eta}}dr^{2}
+ \left ( 1 - \frac{2\eta}{r} \right ) ^{1+\frac{\sigma - 
m}{\eta}}r^{2}d\Omega^{2}
\end{equation}

\begin{equation}
\phi = \frac{\sigma}{2\eta}\ln \left ( 1- \frac{2\eta}{r} \right )
\end{equation}

where $m$ is the mass, $\sigma$ is the scalar charge and $\eta$ is
given by $\eta^{2} = m^{2} + \sigma^{2} $. 
For $\sigma =0$, this solution reduces to the Schwarzschild solution.
Reality of the metric coefficients indicates that we confine ourselves
to the domain $r\ge \eta$. Even for $m= p \eta, \sigma = q\eta$
, with $p,q$ as integers we end up with $p^{2} + q^{2} = 1$ which
contradicts the assumption that $p,q$ be integers. We shall also
see later, that the metric has a naked singularity at $r=2\eta$.
  
We can rewrite the above metric in the Einstein canonical frame
(as written in the papers of Wyman {\cite{wy:81}} and Roberts
{\cite{rob:grg89}}) by employing the standard relations between
the two metrics $g_{\mu\nu}^{str.} = e^{2\phi}g_{\mu\nu}^{E}$. 
Therefore, just as all vacuum solutions of GR are also
solutions of low energy string theory, these scalar field 
solutions are also solutions of the string effective action.
This fact, though trivial, does not seem to have been noticed or mentioned
in the literature 
till now.

Before we get into constructing the charged solutions using the
duality properties of the action and the equations of motion
let us look at the various limiting values of the parameters
$m$,$\sigma$ and $\eta$ in the solution stated above.

For $m=0$ we note that $\eta = \pm \sigma$. 
If we assume $\eta = -\sigma$ then we have :

\begin{equation}
ds^{2} = - \frac{1}{1-\frac{2\eta}{r}} \left (-dt^{2} + dr^{2} \right )
+r^{2}d\Omega^{2}
\end{equation}

On the other hand, with $\eta = +\sigma$ the metric turns out to be :

\begin{equation}
ds^{2} = \left ( 1 - \frac{2
\eta}{r} \right ) ^{2} \left [ \frac{1}{1-\frac{2\eta}{r}} \left ( -dt^{2}
+dr^{2} \right ) + r^{2} d\Omega^{2} \right ] 
\end{equation}

Both these solutions have a  naked singularity at $r=2\eta$. Note also
that the two metrics are conformally related through the factor
$\left (1-\frac{2\eta}{r} \right )^{2}$.

\subsection{\bf Charged Solutions}

Given the above metrics, we can now move on towards constructing the
corresponding electrically charged solutions by using the standard
boosting procedure. This is an application of the target space duality
symmetry of the string effective action (more precisely the set of
transformations which gives the electrically charged solution form a 
subgroup of O(2,1)).

The only new part in the charged metric is in its $g_{00}$ component.
Additionally there does appear changes in the dilaton field and 
a non-zero vector potential appears which actually generates the
electric charge of the solution.

We have :

\begin{equation}
\tilde g_{00} = \frac{g_{00}}{\left [ 1+ (1+g_{00})\sinh^{2} \alpha
\right ] ^{2} } = -\frac{\left ( 1-\frac{2\eta}{r} \right )^{\frac{m+
\sigma}{\eta}}}{\left [ 1+ (1- (1-\frac{2\eta}{r})^{\frac{m+\sigma}{\eta}}
)\sinh^{2} \alpha \right ] ^{2}}
\end{equation}

\begin{equation}
\tilde A_{0} = -
 \frac{(1+g_{00})\sinh 2\alpha}{2\sqrt{2}\left [ 1+ (1+g_{00})\sinh^{2} \alpha
\right ] } = -\frac{(1 - \left ( 1-\frac{2\eta}{r} \right )^{\frac{m+
\sigma}{\eta}}) \sinh 2\alpha}{2\sqrt{2}
\left [ 1+ (1- (1-\frac{2\eta}{r})^{\frac{m+\sigma}{\eta}} \right .
\left .)\sinh^{2} \alpha \right ] }
\end{equation}

\begin{equation}
e^{-2\tilde \phi} = e^{-2\phi} \left [ 1 + (1+ g_{00} ) \sinh^{2} \alpha
\right ] = \left ( 1- \frac{2\eta}{r} \right ) ^{-\frac{\sigma}{\eta}}
\left [ 1 + \left ( 1 - (1-\frac{2\eta}{r})^{\frac{m+\sigma}{\eta}} \right )\sinh^{2}
\alpha \right ] 
\end{equation}

The only nonzero component of the field strength of the Maxwell field
can be shown to be equal to :

\begin{equation}
F_{rt} = \frac{(m+\sigma)\sinh 2\alpha}{r^{2}}\frac{(1-\frac{2\eta}{r})^
{\frac{m+\sigma}{\eta} - 1}}{\sqrt 2 \left [ 1 + (1-(1-\frac{2\eta}{r})
^{\frac{m+\sigma}{\eta}} ) \sinh ^{2}\alpha \right ] ^{2}}
\end{equation}

Therefore, as $r\rightarrow \infty $, $F_{rt} \rightarrow
\frac{(m+ \sigma) \sinh 2\alpha}{\sqrt{2} r^{2}}
$ from which we can read off the expression for the electric charge. 
Also, the mass $M$ can be obtained by comparing the Einstein frame
metric with the Schwarzschild solution. We therefore have,

\begin{equation}
Q_{E} = \frac{(m+\sigma) \sinh 2\alpha}{\sqrt 2} \quad ; \quad
M = m + \left ( m + \sigma \right ) \sinh ^{2} {\alpha}
\end{equation}

Note that for $\sigma = 0$ the scalar field reduces to zero and we
get back the uncharged, Schwarzschild black hole for $\alpha =0$
and the electrically charged stringy black hole for $\alpha \neq 0$.
The presence of a third parameter $\sigma$ is responsible for 
a different relation between $M$ and $Q_{E}$ which can be written as :

\begin{equation}
Q_{E}^{2} = 2 (M-m)(M+\sigma)
\end{equation}

Note, that this relation has a symmetry under the interchange
$m\rightarrow -\sigma ; \sigma \rightarrow -m$.

For $\sigma = 0$ it, ofcourse, reduces to the standard relation
between the charges and masses of the charged black hole. 

If one assumes $m=0$ straightaway we get :

\centerline{${ Q_{E}^{2} 
= 2M (M+ \sigma)} $}

It is easy to see that,
, holding $\sigma \sinh^{2}\alpha$ fixed and setting $\sigma
\rightarrow 0$, $ \alpha \rightarrow \infty$ we obtain :

\begin{equation}
Q_{E}^{2} = 2M^{2} 
\end{equation}

which may be interpreted as the minimum (maximum) electric charge for
a given positive (negative) scalar charge. This limit is quite
similar to the usual {\em extremal} limit one talks about 
in the electric/magnetic solutions. The metric for the 
electric solution in this limit takes the standard form :

\begin{equation}
ds^{2} = - \left ( 1 + \frac{2M}{r} \right )^{-2} dt^{2} +
dr^{2} + r^{2} d\Omega^{2}
\end{equation}

which represents a solution with a null naked singularity at $r=0$.
The naked singularity at $r=2\eta$ has shifted to $r=0$ because we
have taken the $\sigma=\eta\rightarrow 0$ limit.

As is done for the electrically charged black holes, 
we can now use the electric--magnetic
duality symmetry to obtain the corresponding magnetically charged solution
in the string frame.

The magnetically charged solution is related to the electrically
charged one through the relation :

\begin{equation}
ds^{2}_{mag.} = \left ( \exp{(-4\tilde \phi)}\right ) ds^{2}_{elec}   
\end{equation}

which leads to :

\begin{eqnarray}
ds^{2}_{mag.} = 
-\left ( 1 - \frac{2\eta}{r} \right )^{\frac{m-\sigma}{\eta}}
dt^{2} + 
\left ( 1 - \frac{2\eta}{r} \right )^{\frac{-m-\sigma}{\eta}}
\left ( 1 +\left \{ \left ( 1 -\left ( 1 - \frac{2\eta}{r} \right )^{\frac{m+\sigma}{\eta}}\right )
\sinh^{2} \alpha \right \}\right )^{2}  dr^{2} \nonumber \\
+ \left ( 1 - \frac{2\eta}{r} \right )^{1 - \frac{m+\sigma}{\eta}}
r^{2} \left ( 1+ \left \{ \left ( 1 -\left ( 1 - \frac{2\eta}{r} \right )
^{\frac{m+\sigma}{\eta}}\right )
\sinh^{2} \alpha \right \} \right )^{2} d\Omega^{2}
\end{eqnarray}

with the magnetic field given by $ F_{\theta\phi} = Q_{E}\sin \theta
d\theta \wedge d\phi$ and the new dilaton related to the old one
only through a change of sign.

For $\sigma = 0$ and $m=\eta$ we can see that the solution reduces
to the standard magnetically charged black hole with the following
redefinitions :

$ M = \eta \cosh^{2}{\alpha} \quad; \quad Q = \sqrt 2 \eta \sinh \alpha
\cosh \alpha \quad ; \quad  \bar r = r + 2\eta \sinh ^{2} \alpha $

The metric takes the usual form :

\begin{equation}
ds^{2} = \frac{1-\frac{2M}{\bar r}}{1-\frac{Q^{2}}{M\bar r}}dt^{2} + 
\frac{d\bar r^{2}}{\left ( 1-\frac{2M}{\bar r} \right )\left ( 1 -
\frac{Q^{2}}{M\bar r} \right )} + r^{2} d\Omega^{2} 
\end{equation}
 
The other two limits $m =0, \eta = \sigma$ and $m=0, \eta = - \sigma$ turn out 
to yield the
following two metrics :

(1) $ m=0 \quad ,\quad  \eta = \sigma$ :

\begin{equation}
ds^{2} = -\left ( 1 - \frac{2\eta}{r} \right ) ^{-1} dt^{2} + 
\left ( 1+ \frac{2\eta}{r}\sinh^{2}\alpha \right ) \left [ 
\frac{dr^{2}}{1-\frac{2\eta}{r}} + r^{2}d\Omega^{2}\right ] 
\end{equation}

(2) $m=0 \quad , \quad \eta = -\sigma$ : 

\begin{equation}
ds^{2} = - \left (1-\frac{2\eta}{r} \right ) dt^{2} + \frac{\left ( 1-
\frac{2\eta}{r}\cosh^{2} \alpha \right )^{2}}{1-\frac{2\eta}{r}} dr^{2}
+ \left ( r - 2\eta \sinh^{2} \alpha \right )^{2} d\Omega^{2}
\end{equation}

Lastly, setting $m=0$ and then allowing $\sigma \rightarrow 0
$, $\alpha \rightarrow \infty$ with $M = \sigma \sinh^{\alpha}$
fixed yields the usual non--singular infinite throat solution 
given as :

\begin{equation}
ds^{2} = -dt^{2} + \frac{dr^{2}}{\left ( 1 - \frac{2M}{r} \right )^{2}}
+ r^{2} d\Omega^{2} 
\end{equation}

This is the sort of {\em extremal} limit we obtain for this class
of solutions. The only difference of this solution (as well as
the extremal limit of the electrically charged hole) with the
extremal limits of the solutions obtained by boosting the
Schwarzschild  is that
$m$ is now replaced by $\sigma$ and therefore $M=\sigma \sinh^{2}{\alpha}$
The various solutions and their limiting cases are tabulated
in Table I.

Therefore, these solutions can be viewed as generalisations of the
black holes due to Garfinkle, Horowitz and Strominger.
The GHS construction starts off with a Schwarzschild solution as the
seed and
uses the T-duality transformation. On the other hand, if one starts
with a solution in Einstein--massless scalar theory (which
represents a naked singularity) the duality transformations provide
a whole class of nakedly singular solutions which violate the
tenets of cosmic censorship within the scope of low
energy, effective string theory. Setting the dilaton charge to zero
we get back the black holes with regular event horizons. One might
be tempted to say, at this point that the presence of the dilaton,
in a way, forbids the existence of regular horizons in this class of
spacetimes and thereby acts as a parameter which controls the 
existence--non-existence of naked singularities (for previous work
on cosmic censorship and the dilaton see {\cite{hh:prd93}})!

A further generalisation of this class would be to look at 
the possibility of Kerr-type solutions in the Einstein--scalar system
and thereafter utilise the duality symmetries to construct the
corresponding analogs of the Kerr-Sen black holes 
{\cite{sen:rbh}} with the
dilatonic Kerr solution as the {\em seed} metric. Indeed, there is
some work on Kerr--type metrics in Brans--Dicke theory {\cite{ma:ijtp}}
which can be used as the starting point of such a construction.
Recall that low energy, effective string theory in the string frame 
is essentially
Brans--Dicke theory with the $\omega$ parameter being equal to
$-1$. Therefore, such solutions with rotation and a scalar field (dilaton)
can be written down easily by inserting the appropriate value of $\omega$.
Furthermore, using duality transformations one can obtain the 
corresponding charged versions. These cases will be discussed
in a separate article {\cite{sk:prep}}.

\subsection{\bf Solutions using the Buscher formulae}

One can also obtain new solutions by using the Buscher formulae.
For the case when the antisymmetric tensor field is zero and
with no Maxwell field present, we have the simple relations :

\begin{equation}
\bar g_{00} = \frac{1}{g_{00}} \quad ; \quad \bar \phi = \phi -
\frac{1}{2}\ln (-g_{00})
\end{equation}

where $(\bar g_{\mu\nu},\bar \phi )$ is the new solution.

For the string frame metric given earlier, we therefore have 
the new solution given as :

\begin{equation}
ds_{str.}^{2} = -\left ( 1 - \frac{2\eta}{r} \right )^\frac{-m- \sigma}{\eta}
dt^{2} + \left ( 1 - \frac{2\eta}{r} \right )^{\frac{\sigma - m}{\eta}}dr^{2}
+ \left ( 1 - \frac{2\eta}{r} \right ) ^{1+\frac{\sigma - 
m}{\eta}}r^{2}d\Omega^{2}
\end{equation}

with the dilaton as :

\begin{equation}
\bar \phi = -\frac{m}{2\eta}\ln \left ( 1 - \frac{2\eta}{r} \right )
\end{equation}

Notice that this solution goes over to the Schwarzschild (negative or
positive mass  depending on the sign of $\eta$) for $m=0$.
For $\sigma = 0$ we end up with a metric given as :

\begin{equation}
ds^{2} = \left ( 1 - \frac{2\eta}{r} \right )^{\mp 1} \left [ -dt^{2}  
+ dr^{2} + r(r-2\eta)d\Omega^{2} \right ]
\end{equation}

with $m = \pm \eta$. Note that these two solutions are exactly the
same as the ones obtained earlier with $m =0, \sigma = \pm \eta$.
The Buscher transformations, in a sense, interchange the roles of 
$m$ and $\sigma$. The general metrics (for arbitrary $m$ and $\sigma$
) are also mapped onto each other under the interchange :

\centerline{$ m \rightarrow -\sigma \quad ; \quad \sigma \rightarrow
-m $}

Additionally, one should note that one can use the $O(2,1)$ transformation
and the electric--magnetic duality symmetry to obtain newer
solutions from the one obtained above. In particular, all the
previously stated charged solutions will be mapped onto newer
ones by the simple use of the correspondence between $m$ and $\sigma$.

\section{\bf Properties of the solutions}       

We shall now embark upon enumerating the various properties of each of
these solutions in somewhat more detail. More precisely, we focus on the
the kind of singularities present 
and analyse the role of the {\em energy
conditions} (i.e. their violation/conservation) . 
A list of all the Riemann tensor components, Ricci tensors and the
Ricci scalar for a class
of metrics which contains the metrics under discussion here as special
cases is given in the Appendix. We shall make use of these results 
in this section.

\subsection{\bf Singularities}

The Riemann tensor components, after substitution of the various
functional forms for $f(r)$, $g(r)$ and $h(r)$ in the expressions
given in the Appendix turn out to have an inverse dependence on 
the quantity $(r-2\eta)$. This indicates a divergence at $r= 2\eta$.
Note also that the geometry has a horizon at $r=2\eta$ (by virtue 
of the fact that a zero in $g_{00}$ indicates a horizon in the geometry
for static, spheri--symmetric metrics). Thus, we have a spacetime which
virtually ends at $r=2\eta$ where it has a singular horizon. It is easy 
to see that the Riemann tensor components 
for the charged solutions also possess a similar divergence at $r=2\eta$
and therefore constitute solutions with naked singularities.  

\subsection{\bf Energy Conditions}

The origins of the Energy Conditions lie in the right hand
side of the Raychaudhuri equation. The quantity $R_{\mu\nu}\xi^{\mu}
\xi^{\nu}$ is related to the energy momentum tensor through the
use of the Einstein field equations. The imposition of an 
energy condition essentially implies that a geodesic congruence
would focus to a point within a finite value of the affine
parameter--this is the standard {\em focusing theorem}.

We shall be concerned with the Null Energy Condition (NEC) and
the Averaged Null Energy Condition (ANEC) for the string frame
metrics and matter fields given above. 

The NEC and ANEC are stated as follows :

{\bf NEC :} For all null $k^{\mu}$ we must have $T_{\mu\nu}k^{\mu}
k^{\nu} \ge 0$.
In the case of a diagonal $T_{\mu\nu}$ with
components $(\rho, \tau, p, p)$ we therefore need to prove :

\begin{equation}
\rho + \tau \ge 0 \quad ; \quad \rho + p \ge 0
\end{equation}

Physically, the NEC implies the positivity of matter energy density
in all frames of reference. It should be noted that there are
other local Energy Conditions (such as the Weak Energy Condition, 
the Strong Energy Condition and the Dominant Energy Condition
) which also appear as assumptions in the proof of the Hawking--Penrose
singularity theorems. We choose the NEC because it is one
of the weakest amongst all.

{\bf ANEC :} For all null $k^{\mu}$ we must have :

\begin{equation}
\int_{C} T_{\mu\nu}k^{\mu}k^{\nu} d\lambda \ge 0
\end{equation}

where the integration is along a null curve (denoted by
$C$) with 
the parameter $\lambda$ being the generalised affine parameter
in the sense of Hawking and Ellis{\cite{he:book}}).
For null geodesics, $\lambda$ is the usual affine parameter.

It is important to note that for radial null geodesics, the integrand
will contain $\rho +\tau$ only, apart from the usual
exponential of the combination of the dilaton and the redshift function.

As for the local conditions there also exist global
versions of the Weak Energy Condition where the null
curve along which the integration is performed in the
ANEC are replaced by a non--spacelike one. 

The global energy conditions constitute valid assumptions
under which the singularity theorems can be proved.
 
We now discuss each of the cases separately :

{\bf Case 1 : Pure dilatonic solution and its Buscher dual}

The NEC here reduces to the following two inequalities :

{\underline{$\rho + \tau \ge 0$ Inequality}}

\begin{equation}
\left ( 1-\frac{2\eta}{r} \right )^{\frac{m+\sigma}{\eta} - 2} \left [
\frac{4\sigma \left (r-\eta+\sigma\right )}{r^{4}}
\right ] \ge 0
\end{equation}

{\underline{$\rho + p \ge 0 $ Inequality :}}

\begin{equation}
\left ( 1-\frac{2\eta}{r} \right )^{\frac{m-\sigma}{\eta} - 2} \left [
\frac{2\sigma \left (-r+2m+\eta\right )}{r^{4}}
\right ] \ge 0
\end{equation}
 
The first of these holds for all $r\ge 2\eta$ while the second one 
is violated beyond $r=2m+\eta$. For the Buscher dual solution
we make a replacement $m\rightarrow -\sigma,\sigma \rightarrow -m$.
This results in a violation of the first inequality but a conservation
of the second one. One can therefore see the reflection of duality
in the violation/conservation of the NEC. The ANEC evaluated along
radial null geodesics, by virtue of being a global inequality
will hold good for the first solution but will be violated for the
second one.
   
\vspace{.1in}
{{\bf Case 2 : The electric solutions}}
\vspace{.1in}

The NEC inequalities for the electric solutions turn out to be :

{\underline{ $\rho + \tau \ge 0$ Inequality :}}

\begin{equation}
2 \left (1-\frac{2\eta}{r}\right )^\frac{m-\sigma}{r}
\frac{\cosh^{2}{\alpha}\left [ 2\sigma\left(r+\sigma - \eta \right ) \right ]
+\sinh^{2}{\alpha} \left ( 1-\frac{2\eta}{r} \right )^{\frac{m+\sigma}{\eta}}
\left [ 2m \left ( r-m-\eta \right ) \right ] }{r^{2}(r-2\eta)^{2}
\left ( \cosh^{2}\alpha - \left (1-\frac{2\eta}{r}\right )^{\frac{m+\sigma}
{\eta}} \sinh^{2}\alpha
\right )} \ge 0
\end{equation}

{\underline{$\bf \rho + p \ge 0 $ Inequality :}}

\begin{equation}
\frac{\left [ 2\sigma \cosh^{4} \alpha A(r) +
\sinh^{2}\alpha \cosh^{2}{\alpha} \left (1-\frac{2\eta}{r} \right )^\frac{
m+\sigma}{\eta} 
B(r) 
 + 2m \sinh^{4}\alpha \left (1-\frac{2\eta}{r} \right )^{\frac{
2(m+\sigma )}{\eta}}C(r) \right ] } 
{r^{2}(r-2\eta)^{2}
\left ( \cosh^{2}\alpha - \left (1-\frac{2\eta}{r}\right )^{\frac{m+\sigma}
{\eta}} \sinh^{2}\alpha
\right )^{2}
 \left (1-\frac{2\eta}{r}\right )^\frac{\sigma-m}{\eta}}
\ge 0
\end{equation}

where,

$ A(r) = 2m-r+\eta$ ; $B(r) = 2m(4m-r+\eta)+2\sigma (4\sigma + 4m -2\eta
+r)$

$C(r)= r-\eta+2\sigma$

The L.H. S of these two inequalities are shown in Fig. 1(a) and 1(b).
The values of $\sigma, m $ and hence $\eta$ are as follows :
(a) $\sigma = 2m = 1, \eta = \frac{\sqrt{5}}{2}$ (dashed curve)
(b) $m=2\sigma = 1
,
\eta = \frac{sqrt{5}}{2}$ (dot--dashed curve) (c) $\sigma =m = \frac{1}{\sqrt{2}}, \eta = 1$ (solid curve). We choose $\cosh^{2}\alpha = 2$. 
Notice that the $\rho +\tau \ge 0$ inequality is always satisfied
while the second inequality is certainly violated beyond a certain
value of $r>2\eta$. Therefore the local NEC is essentially violated
beyond a certain value of $r>2\eta$.

\vspace{.1in}
{\bf Case 3 : The magnetic solutions}
\vspace{.1in}

The NEC inequalities are given as :

{\underline{$\rho +\tau \ge 0$ Inequality :}}

\begin{equation}
\frac{2\left [-\sigma \cosh^{2}\alpha X(r) +\cosh^{2}\alpha
\sinh^{2}{\alpha} \left (1-\frac{2\eta}{r}\right )^{\frac{m+\sigma}{\eta}}
Y(r) + \sinh^{4}\alpha \left (1-\frac{2\eta}{r}\right )^{\frac{2(m+\sigma)}
{\eta}}
Z(r) \right ]} 
{r^{3} \left (1-\frac{2\eta}{r}\right )^
{2}
\left ( \cosh^{2}\alpha - \left (1-\frac{2\eta}{r}\right )^{\frac{m+\sigma}
{\eta}} \sinh^{2}\alpha
\right )^{2}} \ge 0
\end{equation}

where

$X(r)=1-\frac{\eta+\sigma}{r}$, $ Y(r)=2\sigma\left (1-\frac{\eta+\sigma}{r}
\right )+(m+\sigma)\left (-1 + \frac{3\sigma+m+\eta}{r} \right )$

$Z(r)= -\sigma\left (1-\frac{\eta+\sigma}{r}\right ) -(m+\sigma)
\left (-1+\frac{\sigma-m+\eta}{r} \right )$

{\underline{$ \rho + p \ge 0$ Inequality :}}

\begin{eqnarray}
\frac{\left [-\sigma \cosh^{2}\alpha X(r) +\cosh^{2}\alpha
\sinh^{2}{\alpha} \left (1-\frac{2\eta}{r}\right )^{\frac{m+\sigma}{\eta}}
Y(r) + \sinh^{4}\alpha \left (1-\frac{2\eta}{r}\right )^{\frac{2(m+\sigma)}
{\eta}}
Z(r) \right ]} 
{r^{3} \left (1-\frac{2\eta}{r}\right )^
{2}
\left ( \cosh^{2}\alpha - \left (1-\frac{2\eta}{r}\right )^{\frac{m+\sigma}
{\eta}} \sinh^{2}\alpha
\right )^{2}} \nonumber \\
+\frac{\left(1-\frac
{2\eta}{r}\right )^{\frac{m+\sigma}{\eta}}
\left [r(r-2\eta)-\left ( r-m-\sigma-\eta+r(r-2\eta)P(r)\right )^{2}
\right ]}{r^{2}(r-2\eta)^{2}\left (\cosh^{2}\alpha-\left (1-\frac{2\eta 
 }{r} \right )^{\frac{m+\sigma}{\eta}}\sinh^{2}\alpha \right )^{2}}
\nonumber \\  
(m-\sigma)\frac{2(m-r+\eta)+r(r-2\eta)P(r)}{r^{2}(r-2\eta)^{2}} \ge 0
\end{eqnarray}

where 
$P(r) = -\frac{2(m+\sigma)\left (1-\frac{2\eta}{r}\right )^{\frac{m+\sigma}
{\eta}-1}\sinh^{2}\alpha}{\cosh^{2}\alpha - \left (1-\frac{2\eta}{r}\right
)^{\frac{m+\sigma}{\eta}}\sinh^{2}\alpha} $

These L.H.S of the inequalities are plotted in Figures 2 (a), 2 (b) and
 2 (c) respectively. Fig 2(b) and 2(c) plot the same quantity but
 in different domains of $r$.
The values of $\sigma, m $, $\eta$ and $\cosh^{2}\alpha$ 
are the same as before. 

It is quite clear from the figures that the first inequality
is violated everywhere--infact the violation is infinite
in value as one approaches the point $r=2\eta$. Therefore, there
will also be a violation in the ANEC along radial null geodesics.
The second inequality is satisfied beyond a value $r=r_{0}\ge 2\eta$.
Note that the behaviour of the two inequalities is somewhat 
opposite to those for the electric solutions. Recall that for the
electric solutions the first inequality is satisfied for values of
$r\ge 2\eta$ whereas the second one 
is violated in a similar domain ($r>2\eta$). In the magnetic case,
the first inequality is violated for all $r\ge 2\eta$ while the
second one is conserved beyond a certain $r>2\eta$.  
Therefore, it is worth noting once again that dual solutions
do exhibit a duality in the behaviour of the energy condition
inequalities. In particular a duality is clearly seen for the
ANEC along radial null geodesics. These facts substantiate the
claim made earlier for black holes {\cite{sk:prd97}}
and cosmologies {\cite{sk:plb97}} that there is a duality in the
conservation/violation of the energy conditions for stringy spacetime
geometries.
 
\section{\bf Conclusions}

We first summarize the results obtained. 

(i) Beginning with the known solutions of the Einstein--scalar
system we first construct a variety of solutions by utilising
the duality properties of the theory. Specifically, we obtain
the T-dual partner of the well-known JNW solution, the 
electric and magnetically charged metrics are also obtained
by using the electric-magnetic duality symmetry. The relations
between various charges and masses are also written down.

(ii) Apart from the solutions obtained we analyse the status
of the Null Energy Condition and find it to be violated
in most cases. However, the behaviour of the local inequalities
exhibit a sort of `approximate duality' symmetry. Additionally, 
for the ANEC evaluated
along radial null geodesics we find a clear evidence of
conservation for the electric solution and a violation
for the magnetic solution. Thus, as has been discussed
before in the stringy black hole context{\cite{sk:prd97}}
as well as for  string cosmologies{\cite{sk:plb97}}
we discover a `duality' in the conservation/
violation of the ANEC for the whole class of metrics
parametrized by three quantities $Q$,$M$,$\sigma$. It is
interesting that duality has it's reflection on the
behaviour of the matter fields of the theory--more
precisely on the energy--momentum tensor. In a classical
world, negative energy densities are in a way meaningless
and therefore NEC violating solutions (metrics) should
be ruled out. However, quantum expectation values of the
energy momentum tensor can violate the NEC/ANEC and it is
often said that energy--condition violating solutions
belong to the semi--classical extrapolation of the
classical theory. It must be mentioned though that
exact metrics which solve the semiclassical Einstein
equations are indeed quite rare (for a recent reference
in the context of wormholes see {\cite{hv:prl97}}.      

(iii) The solutions which satisfy the ANEC seem to be a
very new class of spacetimes -- such examples are not
abundant in the literature--and seem to suggest an
extension of the Cosmic Censorship Conjecture to
accomodate spacetimes which satisfy such inequalities.

It remains to be seen whether the reflection of the 
duality symmetries of string theory in the 
energy conditions or their averaged versions is
a generic feature of stringy spacetimes--whether they
be black holes, cosmologies or naked singularities.
The first two have been discussed earlier--this paper
concludes the sequence of examples by dealing with
the case of naked singularities as well.  
Work towards an understanding of a general proof of
the relation between energy conditions and the 
duality symmetries of string theory is in progress and will
be communicated in due course.

\section*{\bf ACKNOWLEDGEMENTS}
The author wishes to thank Sukanta Bose for a careful reading
of the manuscript. Financial support from the Inter--University
Centre for Astronomy and Astrophysics is also gratefully
acknowledged.

\newpage
\appendix

\centerline{\bf APPENDIX}

In this appendix, we list the various Riemann, Ricci tensors and the 
Ricci scalar for a general class of metrics. These are used in various
parts of the paper.

The line element we consider is given as :

\begin{equation}
ds^{2} = -f^{2}(r)dt^{2} + g^{2}(r)dr^{2} + h^{2}(r)d\Omega^{2}
\end{equation}

We choose the one--form basis (static observer's basis) to evaluate
these quantities. This is given as :

\begin{equation}
e^{0} = f(r)dt \quad ; \quad e^{1} = g(r)dr \quad ; \quad 
e^{2} = h(r)d\theta \quad ; \quad e^{3} = h(r) \sin \theta d\phi
\end{equation}

The nonzero components of the Riemann tensor are given as :

\begin{equation}
R^{0}_{110} = \frac{1}{g^{2}} \left ( \frac{f''}{f} - \frac{f'}{f}
\frac{g'}{g} \right ) \quad ; \quad R^{0}_{220} = R^{0}_{330}
=\frac{1}{g^{2}} \frac{f'}{f}\frac{h'}{h} 
\end{equation}

\begin{equation}
R^{2}_{112} = R^{3}_{113} = \frac{1}{g^{2}}\left ( \frac{h''}{h} -
\frac{h'}{h}\frac{g'}{g} \right ) \quad ; \quad 
R^{3}_{232} = \frac{1}{h^{2}} \left ( 1 - \left (\frac{h'}{g} \right )
^{2} \right ) 
\end{equation}

The Ricci tensor components are given as :

\begin{equation}
R_{00} = \frac{1}{g^{2}} \left ( \frac{f''}{f} - \frac{f'}{f}
\frac{g'}{g}  
+ 2  \frac{f'}{f}\frac{h'}{h} \right )
\end{equation}

\begin{equation}
R_{11} = -\frac{1}{g^{2}} \left ( \frac{f''}{f} - \frac{f'}{f}
\frac{g'}{g}  
\right )
 -\frac{2}{g^{2}} \left ( \frac{h''}{h} - \frac{h'}{h}
\frac{g'}{g}  
\right )
\end{equation}

\begin{equation}
R_{22} = R_{33} =  
-\frac{1}{g^{2}} \left ( \frac{h''}{h} - \frac{h'}{h}
\frac{g'}{g}  
+  \frac{f'}{f}\frac{h'}{h} \right )
+ \frac{1}{h^{2}} \left ( 1 - \left (\frac{h'}{g}\right )^{2} \right )
\end{equation}

The Ricci scalar is given as :

\begin{equation}
R = 
 -\frac{2}{g^{2}} \left ( \frac{f''}{f} - \frac{f'}{f}
\frac{g'}{g}  
\right )
 -\frac{4}{g^{2}} \left ( \frac{h''}{h} - \frac{h'}{h}
\frac{g'}{g}  
\right ) - \frac{4}{g^{2}} \frac{f'}{f}\frac{h'}{h}
+ \frac{2}{h^{2}} \left ( 1 - \left (\frac{h'}{g} \right )^{2} \right )
\end{equation}

Given these expressions for $R_{\mu\nu}$ one can then evaluate the
quantitites $\rho+\tau = R_{00}+R_{11}$ and $\rho + p = R_{00}+R_{22}$
which comprise the L.H.S. of the null energy condition inequalities.

\newpage

\centerline{\bf TABLE I}

\begin{center}
\begin{tabular}{|c|c|c|c|c|}\hline
{\bf m} & {\bf $\sigma$} & {\bf $\eta $} & {\bf $\alpha $} & 
{\bf Solution} \\ \hline
Nonzero & Nonzero & Nonzero & 0 & JNW (Wyman) \\ \hline
Nonzero & 0 & $\eta = \pm m$ & 0 & Schwarzschild \\ \hline
0 & Nonzero & $\eta = \pm \sigma $ & 0 & $m= 0$ JNW (Wyman) \\ \hline
Nonzero & Nonzero & Nonzero & Nonzero & (a) Naked Elec. JNW \\ 
& & & & (b) Dual Mag. JNW \\ \hline
0 & Nonzero & $\eta = \pm \sigma $ & Nonzero & (a)$m=0$ Elec. JNW \\ 
& & & & (b) $m=0$ Dual Mag. JNW \\ \hline
Nonzero & 0 & $\eta = \pm m$ & Nonzero & (a) Elec. GHS \\
& & & & (b) Dual Mag. GHS \\ \hline
0 & $\sigma \rightarrow 0, $ 
 & $\eta \rightarrow
0, \frac{\sigma}{\eta}\rightarrow 1$ & $\alpha \rightarrow \infty, 
$ & (a) `Extr.' Elec JNW \\
& $\sigma \sinh^{2} \alpha \rightarrow M$ & & $\sigma \sinh^{2} \alpha
\rightarrow M$ & (b) Dual `Extr.' Mag. JNW \\ \hline
$m \rightarrow 0, $ 
 & 0 & $\eta \rightarrow
0, \frac{m}{\eta}\rightarrow 1$ & $\alpha \rightarrow \infty, 
$ & (a) Extr. Elec GHS \\
$m \cosh^{2} \alpha \rightarrow M $ & & & $m\cosh^{2} \alpha \rightarrow
M$ & (b) Dual Extr. Mag. GHS \\ \hline
\end{tabular}
\end{center}

\vspace{.3in}   

{\centerline{\bf FIGURE CAPTIONS}

{\bf Figure 1(a)} : $\rho +\tau$ versus $r$ for the electric solution. 

{\bf Figure 1(b)} : $\rho+p$ versus $r$ for the electric solution.

{\bf Figure 2(a)} : $\rho +\tau$ versus $r$ for the magnetic solution.

{\bf Figure 2(b)} : $\rho+p$ versus $r$ for the magnetic solution. 

{\bf Figure 2(c)} : $\rho+p$ verus $r$ for the magnetic solution in
a different domain of $r$.
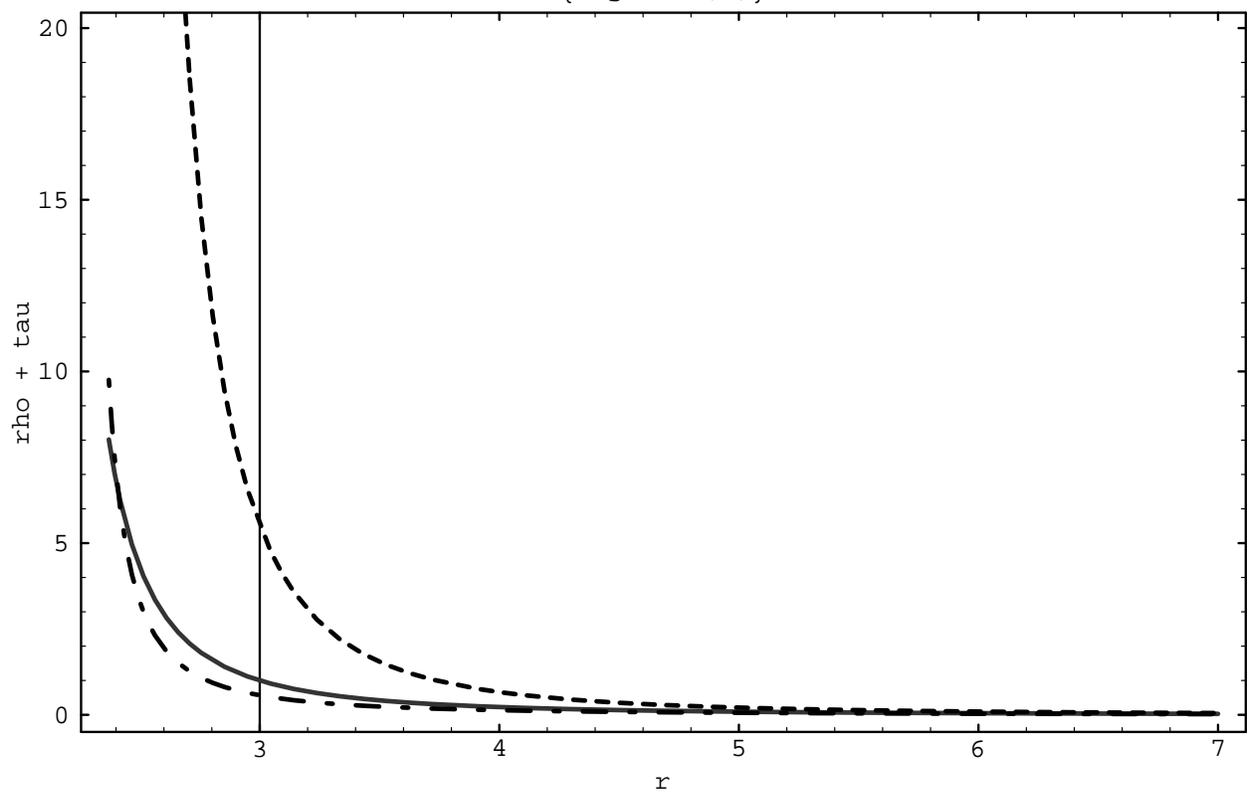
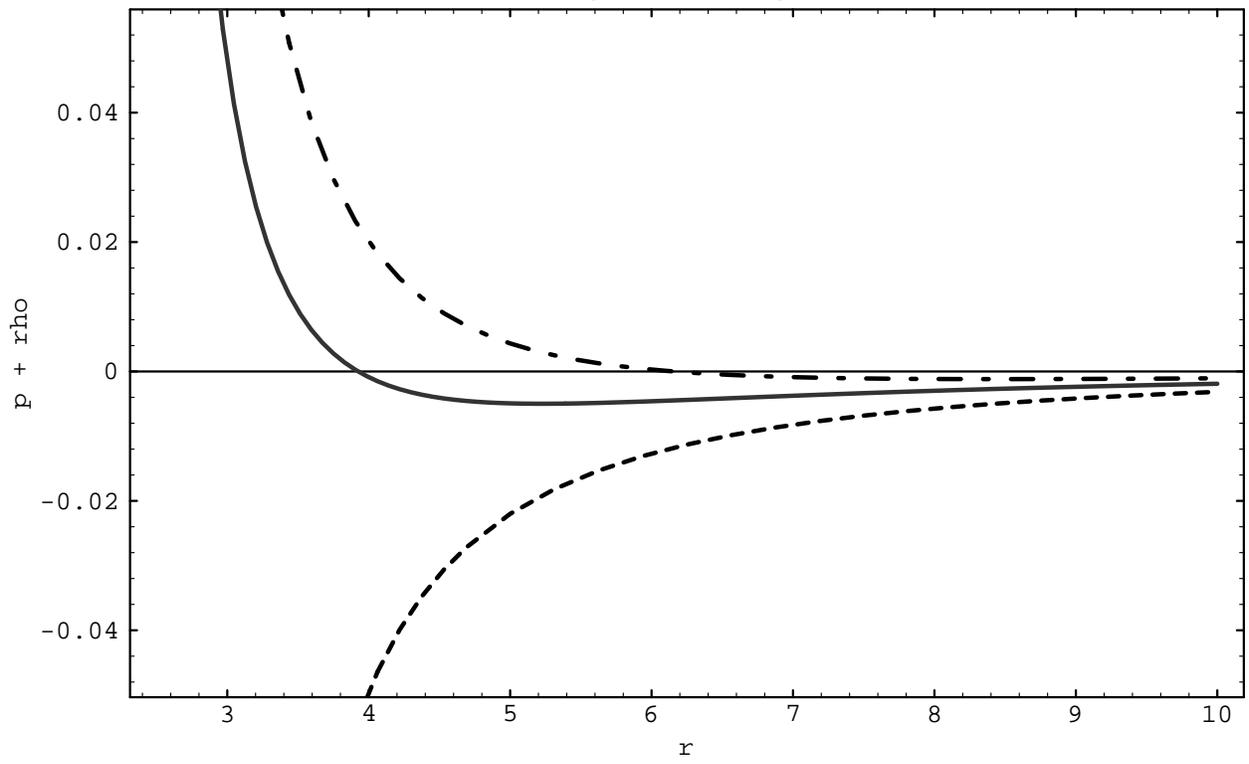
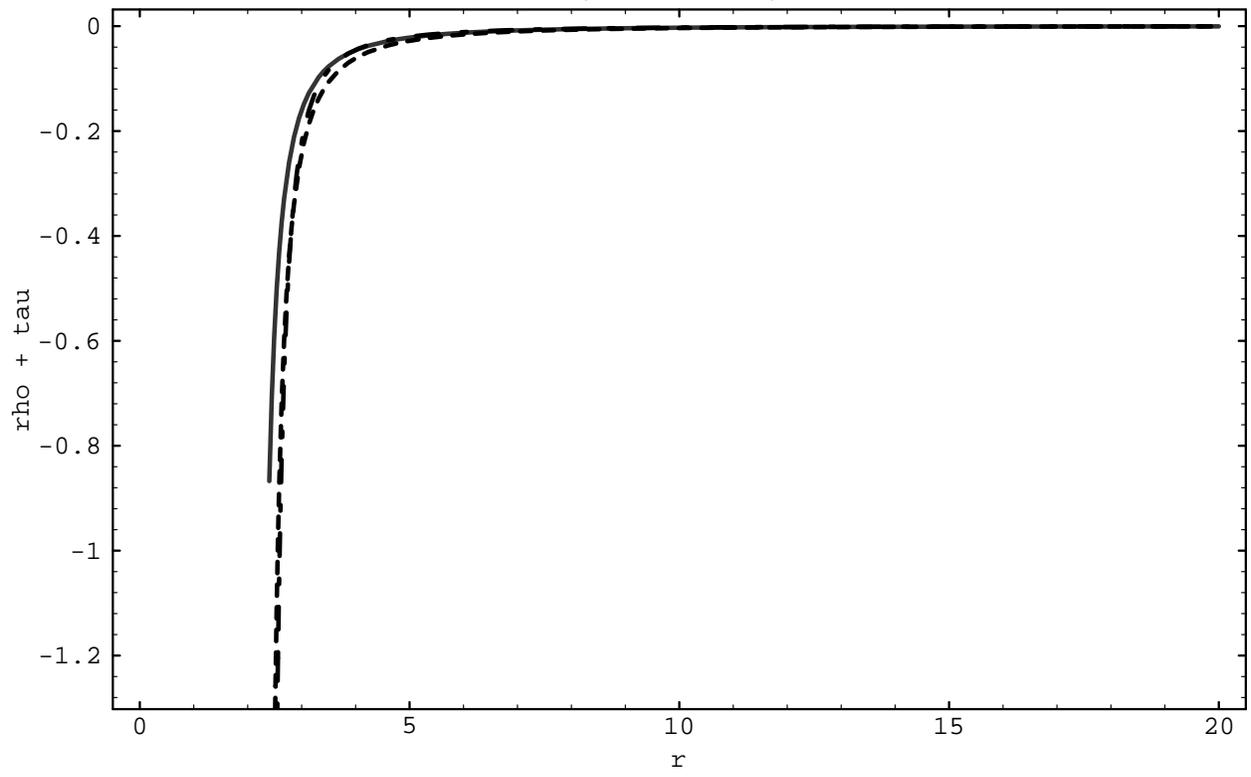
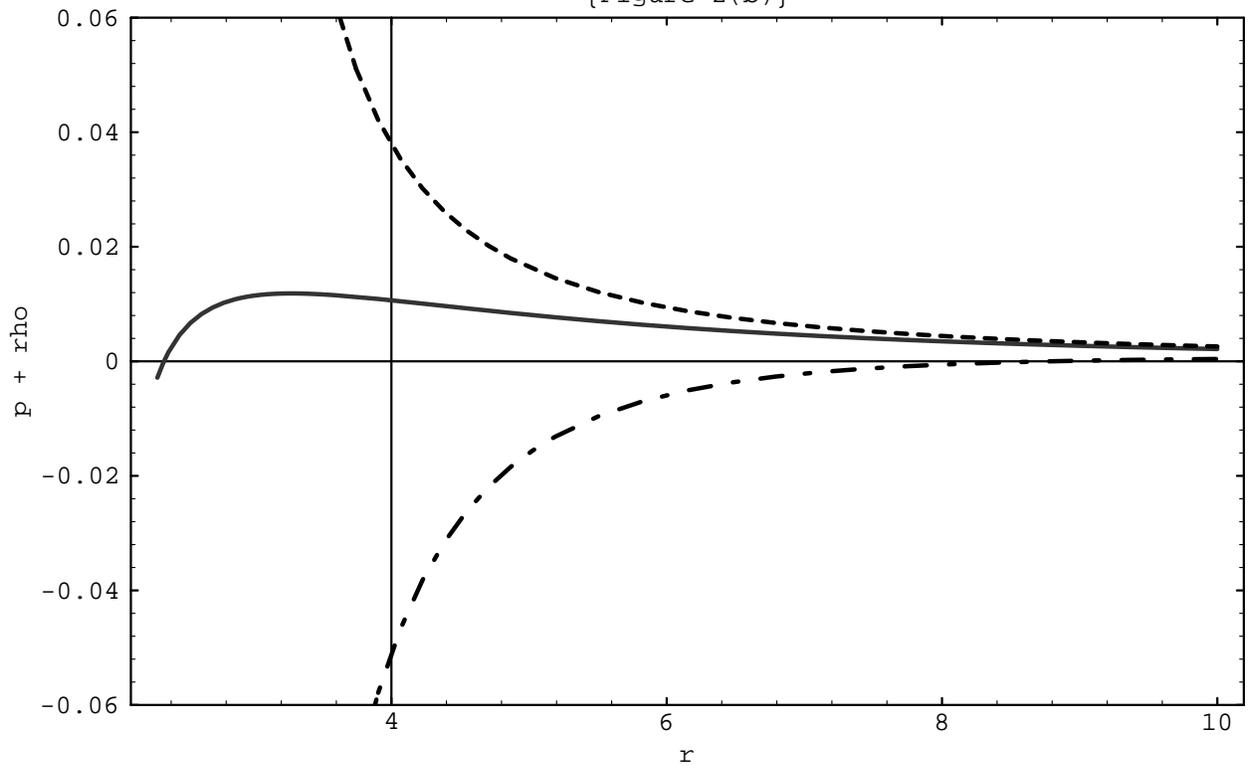
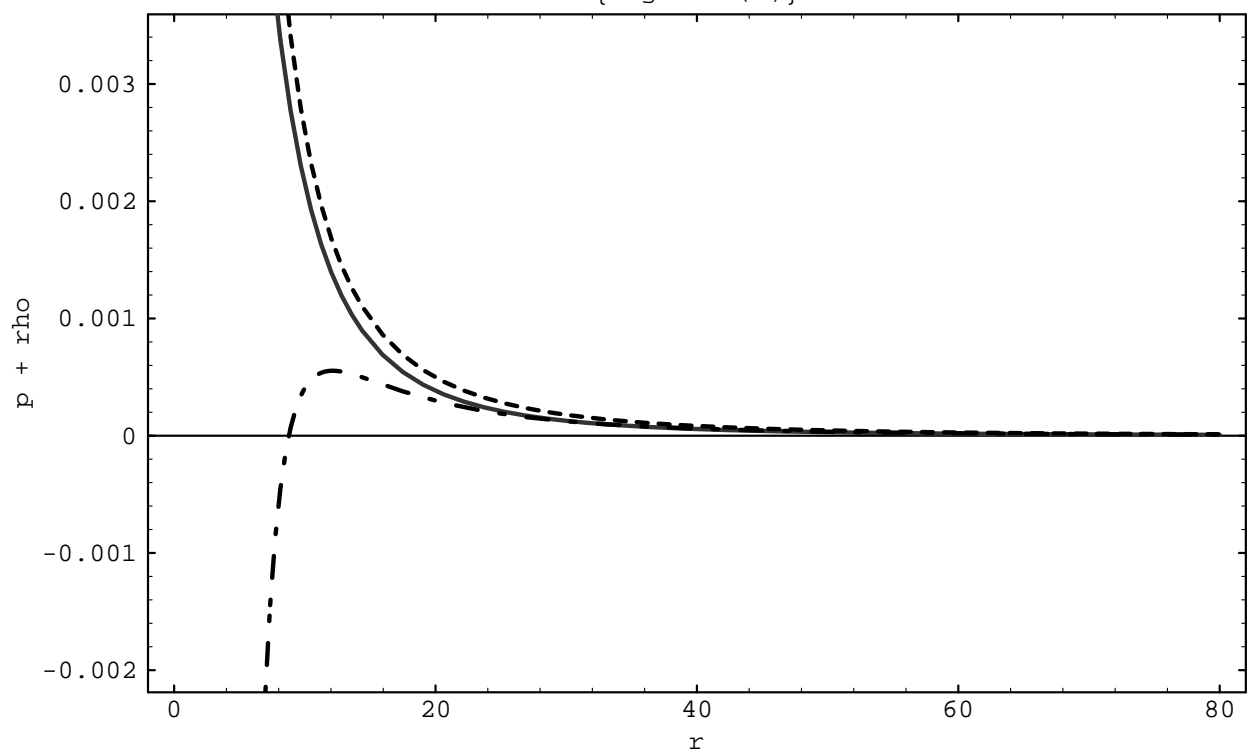
\end{document}